\documentclass{article}

\usepackage{arxiv}

\usepackage[utf8]{inputenc} 
\usepackage[T1]{fontenc}    
\usepackage{hyperref}       
\usepackage{url}            
\usepackage{booktabs}       
\usepackage{amsfonts}       
\usepackage{nicefrac}       
\usepackage{microtype}      
\usepackage{lipsum}

\usepackage{amsmath}
\usepackage{siunitx}
\usepackage{longtable}
\usepackage{array}
\usepackage{multirow}
\usepackage{wrapfig}
\usepackage{float}
\usepackage{colortbl}
\usepackage{pdflscape}
\usepackage{tabu}
\usepackage{threeparttable}
\usepackage{threeparttablex}
\usepackage[normalem]{ulem}
\usepackage{makecell}
\usepackage{xcolor}
\usepackage[round]{natbib}
\title{Optimal multi-wave sampling for regression modelling in two-phase designs}

\author{
  Tong Chen \\
  Department of Statistics\\
  University of Auckland\\
  Email:tche929@aucklanduni.ac.nz\\
   \And
 Thomas Lumley \\
  Department of Statistics\\
  University of Auckland\\
  Email:t.lumley@auckland.ac.nz\\
}

\begin{document}
\maketitle

\begin{abstract}
Two-phase designs involve measuring extra variables on a subset of the cohort where some variables are already measured. The goal of two-phase designs is to choose a subsample of individuals from the cohort and analyse that subsample efficiently. It is of interest to obtain an optimal design that gives the most efficient estimates of regression parameters. In this paper, we propose a multi-wave sampling design to approximate the optimal design for design-based estimators. Influences functions are used to compute the optimal sampling allocations. We propose to use informative priors on regression parameters to derive the wave-1 sampling probabilities because any pre-specified sampling probabilities may be far from optimal and decrease efficiency. Generalised raking is used in statistical analysis. We show that a two-wave sampling with reasonable informative priors will end up with higher precision for the parameter of interest and be close to the underlying optimal design. 
\end{abstract}


\section{Introduction}
\label{sec1}

Large epidemiological studies often collect information on disease status and a large number of covariates for the entire cohort. However, variables of interest, such as risk factors or some expensive exposures, are cost-prohibitive to collect.  It is only possible to measure these variables on a subsample of individuals under a fixed budget. Two-phase stratified sampling \citep{neyman1938contribution} can be useful in this situation. At phase 1, we collect relatively cheap information for the entire cohort, and at phase 2, we sample a small number of individuals from the strata defined by phase-1 data and measure the variables of interest. With considerate choices of stratification and phase-2 sampling probabilities, the two-phase design will result in efficient parameter estimations under a fixed budget constraint \citep{breslow1999design}.

The estimation methods of two-phase designs have been extensively studied, which can be classified into design-based estimation methods and model-based estimation methods. For design-based estimators, weighted likelihood is the most widely used method which weights each observation by its sampling probability. Generalised raking \citep{deville1992calibration,lumley2011connections} improves efficiency by adjusting the sampling probability based on auxiliary variables. For model-based methods, the efficiency gain can be achieved by making assumptions on the outcome model. The maximum likelihood estimator assumes the outcome model is correctly specified, see \cite{scott1997fitting,breslow1997maximum} for discrete phase-1 data and \cite{tao2017efficient} for continuous or discrete phase-1 data. Typically, the maximum likelihood methods are the most efficient estimation methods but not robust to model misspecification \citep{lumley2017robustness}. We focus on the design-based estimation methods in this work.

Compared with estimation methods, the sampling design has not been widely studied. It is of interest to obtain the optimal design, which will include more informative individuals in the phase-2 sample. However, the optimal design will be different for different estimation methods \citep{breslow1999design}. For maximum likelihood estimators, as the outcome model is assumed to be correctly specified, sampling one individual can allow us to extrapolate information about other individuals in the population. For the design-based estimators, it cannot because we do not make any assumptions on the outcome model. Recently, \cite{tao2019} showed that the optimal design for the maximum likelihood estimators would sample from two extreme tails of the derivative of log-likelihood in each stratum when $\beta_x$ is not a strong predictor. This design does not even allow consistent estimations with design-based methods. The optimal design of design-based estimators is Neyman allocation \citep{neyman1934two} applied to the influence functions, which samples relatively evenly across strata.

Optimal design has been considered in some previous works. \citet{reilly1995mean} derived a closed-form expression of the optimal design for their mean-score estimator. Since the expression depends on phase-2 data which are not available at the design stage, \citet{reilly1996optimal} suggested to estimate the expression using data from a further pilot study. \citet{mcisaac2015adaptive} proposed to save the extra cost by using a multi-wave sampling. The idea is to sample wave 1 with pre-specified sampling probabilities and then combine phase-1 and wave-1 data to estimate design components. The later waves can then be sampled adaptively. 

In this work, we exploit an optimal multi-wave sampling approach for design-based estimators. In survey literature, the well-known Neyman allocation \citep{neyman1934two} is the optimal sampling strategy; it minimises the variance of population total for the variable of interest. The regression parameter can be written as the total of its influence functions \citep{breslow2009improved}, so Neyman allocation can then be adopted for minimising the variance of regression parameter. The influence functions also depend on phase-2 data so that a multi-wave sampling can be useful. 

However, wave-1 sampling probabilities and sizes turn out to be important. If the wave-1 sampling probabilities are far from optimal, we will oversample individuals from some less interesting strata. Moreover, a small sample size may lead to the influence functions to be poorly estimated. In this paper, we show that an informative prior on model parameters can improve two-phase design and estimation, even for a non-Bayesian final analysis. The prior information both improves the wave-1 design and regularises the wave-1 analysis. 

The rest of this paper is organised as follows. In Section 2, we define notations and introduce Neyman allocation. In Section 3, the proposed multi-wave sampling and generalised raking are discussed in details. We report results of simulation studies in Section 4. The performance of the proposed sampling method is illustrated using the National Wilms' Tumor Study (NWTS) \citep{d1989treatment,green1998comparison} dataset example in Section 5. Code for all the simulation studies is available from \texttt{https://github.com/T0ngChen/multiwave}.  Remarks are made in Section 6.


\section{Two-phase designs and Neyman allocation}\label{sec2}
\subsection{Notations}
We sample $n$ observations from a cohort of size $N$, where the $i$th observation is sampled with known probability $\pi_i$. Let $Y$ denote the outcome variable, $Z$ denote the inexpensive covariates and $A$ denote the auxiliary variables. We have variables $Z$, $A$ and $Y$ measured for everyone in the cohort at phase 1. Let $X$ be the variable of interest and $X$ is measured on the phase-2 subsample. Let indicator variable $R_i = 1$ if individual $i$ is in the phase-2 sample, otherwise $R_i = 0$, so $E(R_i|Z,A,Y) = \pi_i$. The sampling weight for $i$th observation is $w_i = 1/\pi_i$. 

We refer to $P(Y|X,Z;\beta)$ as the outcome model and $P(X|Z,A;\alpha)$ as the imputation model, so that $Z$ are the components of phase-1 information that we want to put in the outcome model and $A$ are auxiliary variables that are not in the outcome model, but can be used for stratification and imputation.   

In two-phase designs, we assume that the missingness on $X$ only depends on phase-1 data ($P(R|X, Y, Z, A) = P(R|Y, Z, A)$), so that the phase-2 data are missing at random \citep{rubin1976inference}. We use the generalised raking estimator as described in Section \ref{raking} for statistical analysis and our goal is to minimise the variance of $\hat{\beta}_x$ by utilising the optimal multi-wave sampling design.
\subsection{Neyman allocation}\label{sec21}
Suppose the cohort is divided into $H$ strata, and the unbiased estimator of population total for the outcome variable $Y$ can be written as 
\begin{align}
	T_Y = \sum_{h = 1}^H{N_h \Bar{y}_h},
\end{align}
where $\Bar{y}_h$ is the sample mean for stratum $h$. \cite{neyman1934two} derived the optimal sampling allocation to minimise the sampling variance of an estimator of a total with respect to the constraint $n_1 + n_2 + \dots + n_H = n$. It is can be expressed as
\begin{align}
	n_i = \frac{n N_i \sigma_i}{\sum_{h = 1}^H{N_h\sigma_h}}, \label{alloca}
\end{align}
where $\sigma_i$ is the population standard deviation for stratum $i$, $n_i$ and $N_i$ are the phase-2 and phase-1 sample size for stratum $i$ respectively.

However, Neyman allocation is based on the assumption that $n$ is a continuous variable, so the value of $n_i$ calculated from Equation (\ref{alloca}) is not an integer in general. The usual practice is to round to the nearest integer and fiddle around until the sum matches the constraint, but this does not always lead to the optimal allocation.  \citet{wright2017exact} derived an integer-valued algorithm to find an exact optimal allocation; it is equivalent to the Huntington-Hill method used to assign US Congress seats to states.

\section{Multi-wave sampling for design-based estimators}
\label{sec3}
\subsection{Neyman allocation for $\beta_x$}
We are interested in improving the precision for the regression parameter $\beta_x$, so we need to write $\beta_x$ as a total. \citet{breslow2009improved} noted that an estimator of the regression parameter could be written as a total of its influence functions, so we have
\begin{align}
	\sqrt{N}(\hat{\beta} -\beta_0) = \sum_{i=1}^{N}\mathbf{h}_i(\beta) + o_p(N^{-1/2}), \label{infl}
\end{align}
where $h_i(\beta)$ is influence function for observation $i$ in the cohort. It can be approximated by delta-betas which is the change in $\hat{\beta}$ when observation $i$ is deleted. According to Equation (\ref{infl}), a weighted estimator $\beta_w$ can be written as 
\begin{align}
	\sqrt{n}(\hat{\beta}_w -\beta_0) = \sum_{i=1}^{n}w_i\mathbf{h}_i(\beta) + o_p(n^{-1/2}). \label{inflw}
\end{align}

Substituting the standard deviation of influence functions in Equation (\ref{alloca}), we have the optimal continuous allocation
\begin{align}
	n_i = \frac{n N_i \text{Var}(h_i(\beta))^{1/2}}{\sum_{h = 1}^H{N_h \text{Var}(h_h(\beta))^{1/2}}}.\label{neyaa}
\end{align}

This is the same formula as \citet{mcisaac2015adaptive} who derived by directly minimising the estimated variance. In this work, we use the integer-valued algorithm \citep{wright2017exact} to find a global optimal allocation which is more efficient than Neyman.

\subsection{Multi-wave sampling with priors} \label{samplemethod}
The influence functions depend on phase-2 variable $X$, and we do not have any information about it at the design stage. \citet{mcisaac2015adaptive} showed that a multi-wave sampling was helpful. Based on their ideas, the wave 1 can be sampled with pre-specified sampling probabilities and the influence functions can then be estimated.

We can further improve design efficiency by finding a better choice of wave 1. On the one hand, any pre-specified sampling probabilities may be far from optimal, so bad decisions of wave 1 will oversample some less informative individuals. On the other hand, as we want the influence functions $h(\beta)$ but end up with having the estimated influence functions $h(\hat{\beta})$, a relatively small sample size may lead to the influence functions to be poorly estimated.

If we have informative priors on the parameters in both outcome and imputation model, the influence functions can be derived by combining phase-1 data and priors. The optimal wave-1 allocation can be estimated by Equation (\ref{neyaa}). Based on this idea, we propose the following optimal multi-wave sampling design:
\begin{enumerate}
	\item Combine priors, phase-1 data, outcome model and imputation model to compute posterior distributions for $\alpha$, $\beta$, and $X$.
	\item Impute $X$ for all the cohort subjects and estimate the influence functions. 
	\item  Derive the optimal wave-1 sampling allocations using integer-valued Neyman allocation \citep{wright2017exact} and sample wave 1.
	\item  Combine the posterior distributions, wave-1 and phase-1 data to sample wave 2.
\end{enumerate}
The later waves can be sampled adaptively if needed. We also add a constraint $n_h \geq 2$ for wave 1 to ensure a valid variance estimation within each stratum. 

\subsection{Generalised raking}\label{raking}
Generalised raking is a more efficient class of design-based estimators. The idea is to adjust sampling weights using a vector of auxiliary variables $S_i$, and it satisfies the calibration constraints
\begin{equation*}
	\sum_{i \in \operatorname{sample}} g_{i} \frac{R_{i}}{\pi_{i}} S_{i}=\sum_{i \in \operatorname{cohort}} S_{i},
\end{equation*}
where $g_i$ is the calibrated weight. It can be obtained by minimising the total weight change 
\begin{equation*}
	\sum_{i \in \operatorname{sample}} d\left(\frac{g_{i}}{\pi_{i}}, \frac{1}{\pi_{i}}\right),
\end{equation*}
under a given distance measure while satisfying the calibration constraints. \citep{deville1992calibration}. We use Poisson distance $d(a,b)=a\log(a/b)+(b-a)$ in this work. 

In two-phase designs, we are interested in improving the efficiency of regression parameters in the outcome model. The variables in the outcome model cannot be directly used as generalised raking variables because the generalised raking variables should be linearly correlated with regression parameters $\hat{\beta}$, and $X$ and $Y$ are approximately uncorrelated with $h_i$ \citep{lumley2011connections}. According to Equation (\ref{infl}) and (\ref{inflw}), influence functions can be used as generalised raking variables. The efficient design-based estimators use $E[h_i(\beta_0)|A,Y,Z]$ as the auxiliary variable. This is the same class as AIPW \citep{robins1994estimation} estimators \citep{lumley2011connections}.

\citet{kulich2004improving} derived an efficient doubly weighted estimator and proposed a ``plug-in" method to approximate the optimal choice of auxiliary variables. \citet{breslow2009improved,breslow2009using} adopted the ``plug-in" method to conduct imputation generalised raking for case-cohort studies; \citet{rivera2016using} used the same method in the analysis of counter-matched samples. We use the same technique to get the generalised raking variables.   

It is worth to note that, priors are not used in the statistical analysis because currently it is not standard in these fields to do Bayesian analysis.  Arguably even a better option will be to do the Bayesian analysis. We show that we can still gain from design even if we cannot do that.


\section{Simulation Study}\label{sec4}

We conduct extensive simulation studies to evaluate the efficiency of our proposed sampling design. We examine the situation that the exposure of interest is cost-prohibitive, but there exists an inexpensive surrogate variable of it. 

1000 phase-1 samples of size 1000 were simulated. A binary variable of interest $X$ was generated with 15\% exposure, so $X \sim \text{Bern(0.15)}$. A surrogate variable $A$ was simulated with pre-specified sensitivity and specificity. We also simulated a continuous covariate $Z_1 \sim \text{U}(0,1)$ and a binary covariate $Z_2 \sim \text{Bern(0.6)}$. The binary outcome variable $Y$ was simulated using outcome model
\begin{align*}
	Pr(Y|X,Z_1,Z_2) = \text{expit}(\beta_0 + \beta_1 X + \beta_2 Z_1 + \beta_3 Z_2),
\end{align*}
where $\text{expit}(X) = \exp(X)/(1+\exp(X))$, $\beta_0 = -2$ and $\beta_2 = \beta_3 = 1$. The data were divided into $8$ strata based on binary variables $Z_2$, $A$ and $Y$. The imputation model was $P(X|A,Z_1,Z_2)$. We were interested in priors centered either close to or far from the truth with small or moderate variance. Typically, two-wave sampling designs were considered because they were widely used and relatively easy to implement.

We implemented the following designs in our simulation studies:
\begin{enumerate}
	\item A single wave proportional stratified sampling design where phase-2 strata sizes were proportional to phase-1 strata sizes.
	
	\item A single wave balanced stratified sampling design where phase-2 strata sizes were the same.
	
	\item An optimal sampling design where the sampling allocations were derived using the whole data ($X$ also available for individuals not in phase-2 sample). This cannot be done in practice.
	
	\item Two-wave sampling designs where balanced or proportional stratified sampling was used at wave 1.
	
	\item Our proposed two-wave sampling designs (Section \ref{samplemethod}) where informative normal priors on the parameters of outcome model and imputation model were used. $4$ different normal priors were considered, specifically, prior $1$: $\beta_i \sim N(\beta_i-\sqrt{0.1}/2, 0.1)$, $\alpha_j \sim N(\alpha_j-\sqrt{0.1}/2, 0.1)$; prior 2: $\beta_i \sim N(\beta_i-\sqrt{0.1}/2, 1)$, $\alpha_j \sim N(\alpha_j-\sqrt{0.1}/2, 1)$; prior 3: $\beta_i \sim N(\beta_i-1/2, 0.1)$, $\alpha_j \sim N(\alpha_j-1/2, 0.1)$; prior 4: $\beta_i \sim N(\beta_i-1/2, 1)$, $\alpha_j \sim N(\alpha_j-1/2, 1)$.
	
\end{enumerate}
Let $n_a$ be the number of individuals sampled at wave 1. For two-wave sampling designs, we considered 5 different choices for the proportion of phase-2 samples selected at wave 1 ($n_a/n$), which ranged from $1/6$ to $5/6$ in $1/6$ increments. Generalised raking described in Section \ref{raking} was followed. 

Results were presented in terms of Mean Squared Error (MSE) and Empirical Relative Efficiency (ERE) to the optimal design for the parameter of interest $\beta_1$. MSE is the average squared difference between the estimated values $\hat{\beta}_1$ and the actual value $\beta_1$. Larger values of MSE indicate lower efficiency. ERE of design $A$ is defined as the ratio of the empirical variance of $\hat{\beta_1}$ from the optimal design to the empirical variance of $\hat{\beta_1}$ from design $A$ \citep{mcisaac2015adaptive}. Values of ERE smaller than $1$ indicate a loss of efficiency compared with the optimal design.

Results were shown in Table \ref{sim_1}. Two-wave sampling designs were slightly more efficient than optimal design in some settings, these were consistent with simulation studies of \cite{mcisaac2015adaptive}, because the underlying optimal designs are only optimal when sample size goes to infinity. Besides that, Neyman allocation is not the optimal design for the raking estimator. However, the optimal design of raking estimators requires the true influence functions, which are not available in practice. 

In our simulation studies, single wave balanced stratified sampling designs were more efficient than single wave proportional stratified sampling designs, but single wave sampling designs did not achieve near optimality. 

Two-wave sampling designs with pre-specified wave 1 generally performed better than single wave sampling designs except in the situation that wave 1 was small, because influence functions tended to be poorly estimated with small amount of phase-2 data. With the increase of wave-1 sample size, we got more precise estimates of influence functions which would lead to a better design, but the disadvantage was that we also had to sample fewer people at wave 2. When wave-1 size was around half, two-wave designs had a better performance and were more efficient than single wave sampling designs. In addition, wave-1 sampling probabilites also turned out to be essential. Two-wave designs with balanced sampling at wave 1 were more efficient than two-wave designs with proportional sampling at wave 1. 

Our proposed designs were very close to optimal design for the priors centered either close to or far from the truth with small or moderate variance. Table \ref{sim_1} showed that all the four priors resulted in good wave 1 allocations, so a small wave 1 did not lose efficiency here. 

Other simulation studies (not shown here) showed that well-calibrated tight priors performed slightly better than well-calibrated flat priors but poorly-calibrated flat priors performed better than poorly-calibrated tight priors. As weakly informative priors prevented us from getting extreme inference \citep{gelman2008weakly}, we recommended to use weakly informative priors.

\section{Data Example}
We illustrated the performance of our proposed sampling design using data from the National Wilms’ Tumor study (NWTS) \citep{d1989treatment, green1998comparison}. The data consist of $N=3915$ observations. The variables available for all the individuals include histology evaluated by institution (favorable vs. unfavorable (instit)), histology evaluated by central lab (favorable vs. unfavorable (histol)), stage of disease (I-IV (stage)), age at diagnosis (age), diameter of tumor (tumdiam), study (3 vs. 4 (study)) and indicator of relapse (relapse). We assumed central lab histology was only available at phase 2 and was the variable of interest. All the other variables were assumed to be available for the whole cohort.
Following \citep{kulich2004improving,breslow2009using,rivera2016using}, a similar outcome model was fitted as 
\begin{align*}
	Pr(\text{relapse}|&\text{histol,age}_1\text{,age}_2\text{,stage}_1\text{,tumdiam}) = \text{expit}(\beta_0 + \beta_1 \text{histol} + \beta_2 \text{age}_1 \\ &+ \beta_3 \text{age}_2 + \beta_4 \text{stage}_1 + \beta_5 \text{tumdiam} +\beta_6  \text{tumdiam} \times \text{stage}_1) ,
\end{align*}
where age$_1$ and age$_2$ were a linear spline with separate slope for greater or less than 1 year old and stage$_1$ was a binary indicator (III–IV vs. I–II). We took institutional histology as central lab histology measured with error, so it was a good surrogate variable. Central lab histology was imputed using a logistic model with predictors institutional histology, age$_3$ ($>$10
years vs. $<$10 years), stage$_2$ (IV vs. I–III), study and the interaction between study and stage$_2$. 

The data were divided into $8$ strata based on institutional histology, relapse and study with strata sizes $(1257,1769,\\107,113,223,284,84,78)$. In the simulation study, $720$ individuals were sampled at phase 2. Based on above outcome model and the whole cohort data, the optimal phase-2 sample sizes for each stratum were $n_{opt} = (156, 241,  38,  39,  75,\\ 111,  36,  24)$.

$1000$ phase-2 samples were simulated. Similarly to previous simulation studies, we examined five choices of wave-1 sample sizes and implemented a single wave balanced sampling design, a single wave optimal sampling design based on full data, a two-wave sampling design with balanced sampling at wave 1 and our proposed sampling designs. We considered $4$ different normal priors, specifically, prior $1$: $\beta_i \sim N(\beta_i-\sqrt{0.1}/2, 0.1)$, $\alpha_j \sim N(\alpha_j-\sqrt{0.1}/2, 0.1)$; prior 2: $\beta_i \sim N(\beta_i-\sqrt{0.1}/2, 1)$, $\alpha_j \sim N(\alpha_j-\sqrt{0.1}/2, 1)$; prior 3: $\beta_i \sim N(\beta_i-1/2, 0.1)$, $\alpha_j \sim N(\alpha_j-1/2, 0.1)$; prior 4: $\beta_i \sim N(\beta_i-1/2, 1)$, $\alpha_j \sim N(\alpha_j-1/2, 1)$.

Proportional stratified sampling was not considered because the sampling probabilities for each stratum were $(0.32, 0.45,0.03,0.03,0.06,0.07,0.02,0.02)$. If $100$ individuals were sampled at wave 1, it would only sample $2$ individuals from seventh and eighth strata, so the influence functions would be very poorly estimated and the variance of influence functions for these strata might not exist.

Results were presented in Table \ref{nwts_sim}. Single wave balanced stratified sampling design was not close to optimal. Two-wave sampling with balanced stratified sampling at wave 1 performed slightly better but still was not close to optimal for all the choices of wave-1 sample sizes. Our  proposed  designs  were still very  close  to  optimal  design  for all the four priors. As the NWTS data have rich phase-1 information, efficiency gains of our proposed design were also from using the rich phase-1 data at the design stage.

%

\section{Discussion}
\label{disc}
We described a multi-wave adaptive sampling approach to approximate the optimal two-phase design for fitting a regression model using design-based estimators. The prior knowledge of parameters and phase-1 data combine to be usable to approximate the optimal wave 1, so we use the whole cohort information at the design stage even before phase-2 sampling. With reasonable well-calibrated priors, the efficiency of the proposed design is very close to the underlying optimal design.

There are two main advantages of using priors. Firstly, it is obviously that priors help us put in available information. Based on analysing the rich and readily available medical data (e.g., electronic health record), genuine clinical knowledge and previous studies, it is reasonable to have useful prior knowledge. Moreover, even if we do not have much information, weakly informative priors are also found to be useful because they regularise extreme estimations in the wave-1 analysis which occasionally happen using completely non-informative priors or maximum likelihood \citep{gelman2008weakly}.

Single wave sampling designs are not efficient in general. Balanced stratified sampling designs are more efficient than proportional stratified sampling designs, but still do not often achieve near optimality \citep{breslow1999design}. 

\citet{mcisaac2015adaptive} showed a multi-wave adaptive sampling approach for fitting a regression model with the mean-score method. Under stratified sampling, the mean-score method is the same as the weighted likelihood. They derived the optimal design by directly minimising the estimated variance, so it is the same as Neyman allocation with influence functions. 

For the multi-wave sampling design with pre-specified wave 1, there is a trade-off between wave-1 sample size and design efficiency. A larger wave 1 will result in better estimations of wave-1 allocations but less individuals to be sampled at later waves, whereas a smaller wave 1 will lead to poorer estimated influence functions and a worse design. Reasonable informative priors can help us get better estimations of influence functions even with small amount of phase-2 data.

The improvement of design efficiency can be further achieved by changing strata because strata are also pre-specified. As estimates of influence functions can be derived using priors, we are very likely to find a better stratification before phase-2 sampling. This will be discussed in future works.

\section*{ACKNOWLEDGMENTS}
The authors wish to acknowledge the use of New Zealand eScience Infrastructure (NeSI) high performance computing facilities, consulting support and/or training services as part of this research. New Zealand's national facilities are provided by NeSI and funded jointly by NeSI's collaborator institutions and through the Ministry of Business, Innovation \& Employment's Research Infrastructure programme. URL https://www.nesi.org.nz.

This work was supported in part by the University of Auckland doctoral scholarship,
 Patient Centered Outcomes Research Institute
(PCORI) Award R-1609-36207 and U.S. National Institutes of Health (NIH) grant R01-
AI131771. 

\begin{table}[H]
	\caption{Mean squared error (MSE) and empirical relative efficiency (ERE) to the optimal design for $\beta_1$ based on $1000$ Monte Carlo simulations.}
	\centering
	\scalebox{0.73}{\begin{tabular}{lllllllllllllllllll}
			\toprule
			\multicolumn{1}{c}{ }&\multicolumn{1}{c}{ } & \multicolumn{1}{c}{Opt} & \multicolumn{2}{c}{Prior 1} & \multicolumn{2}{c}{Prior 2} & \multicolumn{2}{c}{Prior 3} & \multicolumn{2}{c}{Prior 4} & \multicolumn{2}{c}{Two.prop} & \multicolumn{2}{c}{Two.bal} & \multicolumn{2}{c}{Single.prop} & \multicolumn{2}{c}{Single.bal} \\
			\cmidrule(l{3pt}r{3pt}){3-3} \cmidrule(l{3pt}r{3pt}){4-5} \cmidrule(l{3pt}r{3pt}){6-7} \cmidrule(l{3pt}r{3pt}){8-9} \cmidrule(l{3pt}r{3pt}){10-11} \cmidrule(l{3pt}r{3pt}){12-13} \cmidrule(l{3pt}r{3pt}){14-15} \cmidrule(l{3pt}r{3pt}){16-17} \cmidrule(l{3pt}r{3pt}){18-19}
			$(\beta_1,se,sp)$ & $n_a/n$  & \small{MSE*} & \small{MSE*} & \small{ERE} & \small{MSE*} & \small{ERE} & \small{MSE*} & \small{ERE} & \small{MSE*} & \small{ERE} & \small{MSE*} & \small{ERE} & \small{MSE*} & \small{ERE} & \small{MSE*} & \small{ERE} & \small{MSE*} & \small{ERE}\\
			\midrule
			& 1/6 & 0.74 & 0.79 & 0.94 & 0.80 & 0.93 & 0.82 & 0.91 & 0.77 & 0.96 & 2.32 & 0.32 & 1.54 & 0.49 & 1.07 & 0.70 & 0.88 & 0.85\\
			
			& 2/6 & 0.74 & 0.87 & 0.86 & 0.83 & 0.90 & 0.77 & 0.96 & 0.78 & 0.95 & 1.09 & 0.68 & 0.95 & 0.79 & 1.07 & 0.70 & 0.88 & 0.85\\
			
			& 3/6 & 0.74 & 0.74 & 1.00 & 0.77 & 0.97 & 0.86 & 0.87 & 0.82 & 0.91 & 1.11 & 0.67 & 0.89 & 0.84 & 1.07 & 0.70 & 0.88 & 0.85\\
			
			& 4/6 & 0.74 & 0.81 & 0.92 & 0.77 & 0.96 & 0.84 & 0.88 & 0.87 & 0.85 & 0.92 & 0.80 & 0.82 & 0.90 & 1.07 & 0.70 & 0.88 & 0.85\\
			
			\multirow[t]{-5}{*}{\raggedright\arraybackslash \small{$(0.5,0.8,0.8)$}} & 5/6 & 0.74 & 0.81 & 0.92 & 0.76 & 0.98 & 0.84 & 0.88 & 0.84 & 0.88 & 1.00 & 0.74 & 0.84 & 0.89 & 1.07 & 0.70 & 0.88 & 0.85\\
			\cmidrule{1-19}
			& 1/6 & 0.54 & 0.55 & 0.99 & 0.59 & 0.92 & 0.59 & 0.92 & 0.55 & 0.99 & 1.14 & 0.48 & 0.67 & 0.81 & 0.84 & 0.65 & 0.62 & 0.88\\
			
			& 2/6 & 0.54 & 0.57 & 0.94 & 0.57 & 0.96 & 0.59 & 0.93 & 0.58 & 0.94 & 0.65 & 0.84 & 0.63 & 0.86 & 0.84 & 0.65 & 0.62 & 0.88\\
			
			& 3/6 & 0.54 & 0.58 & 0.94 & 0.58 & 0.94 & 0.54 & 1.00 & 0.56 & 0.96 & 0.65 & 0.83 & 0.59 & 0.93 & 0.84 & 0.65 & 0.62 & 0.88\\
			
			& 4/6 & 0.54 & 0.56 & 0.96 & 0.60 & 0.91 & 0.62 & 0.88 & 0.55 & 0.99 & 0.70 & 0.78 & 0.60 & 0.92 & 0.84 & 0.65 & 0.62 & 0.88\\
			
			\multirow[t]{-5}{*}{\raggedright\arraybackslash \small{$(0.5,0.9,0.9)$}} & 5/6 & 0.54 & 0.58 & 0.93 & 0.56 & 0.97 & 0.62 & 0.87 & 0.60 & 0.90 & 0.73 & 0.74 & 0.58 & 0.94 & 0.84 & 0.65 & 0.62 & 0.88\\
			\cmidrule{1-19}
			& 1/6 & 0.78 & 0.79 & 0.99 & 0.78 & 1.00 & 0.82 & 0.94 & 0.83 & 0.94 & 2.48 & 0.33 & 1.65 & 0.50 & 1.04 & 0.74 & 0.90 & 0.87\\
			
			& 2/6 & 0.78 & 0.81 & 0.96 & 0.77 & 1.01 & 0.89 & 0.87 & 0.77 & 1.01 & 1.19 & 0.66 & 0.98 & 0.82 & 1.04 & 0.74 & 0.90 & 0.87\\
			
			& 3/6 & 0.78 & 0.74 & 1.05 & 0.82 & 0.95 & 0.85 & 0.91 & 0.82 & 0.94 & 1.02 & 0.76 & 0.91 & 0.89 & 1.04 & 0.74 & 0.90 & 0.87\\
			
			& 4/6 & 0.78 & 0.80 & 0.97 & 0.82 & 0.94 & 0.83 & 0.94 & 0.81 & 0.96 & 0.96 & 0.80 & 0.90 & 0.88 & 1.04 & 0.74 & 0.90 & 0.87\\
			
			\multirow[t]{-5}{*}{\raggedright\arraybackslash \small{$(1,0.8,0.8)$}} & 5/6 & 0.78 & 0.83 & 0.94 & 0.80 & 0.97 & 0.78 & 0.99 & 0.84 & 0.92 & 1.04 & 0.75 & 0.89 & 0.87 & 1.04 & 0.74 & 0.90 & 0.87\\
			\cmidrule{1-19}
			& 1/6 & 0.57 & 0.58 & 0.98 & 0.62 & 0.91 & 0.58 & 0.98 & 0.59 & 0.96 & 4.58 & 0.13 & 0.75 & 0.77 & 0.81 & 0.70 & 0.61 & 0.94\\
			
			& 2/6 & 0.57 & 0.59 & 0.96 & 0.55 & 1.03 & 0.59 & 0.96 & 0.58 & 0.98 & 0.78 & 0.75 & 0.61 & 0.93 & 0.81 & 0.70 & 0.61 & 0.94\\
			
			& 3/6 & 0.57 & 0.57 & 1.00 & 0.56 & 1.02 & 0.61 & 0.93 & 0.58 & 0.98 & 0.63 & 0.92 & 0.58 & 1.00 & 0.81 & 0.70 & 0.61 & 0.94\\
			
			& 4/6 & 0.57 & 0.58 & 0.98 & 0.59 & 0.96 & 0.60 & 0.96 & 0.60 & 0.95 & 0.74 & 0.77 & 0.54 & 1.07 & 0.81 & 0.70 & 0.61 & 0.94\\
			
			\multirow[t]{-5}{*}{\raggedright\arraybackslash \small{$(1,0.9,0.9)$}} & 5/6 & 0.57 & 0.60 & 0.94 & 0.58 & 0.98 & 0.61 & 0.93 & 0.59 & 0.97 & 0.76 & 0.75 & 0.62 & 0.92 & 0.81 & 0.70 & 0.61 & 0.94\\
			\cmidrule{1-19}
			& 1/6 & 0.88 & 0.93 & 0.95 & 0.89 & 0.98 & 0.98 & 0.90 & 0.92 & 0.95 & 9.69 & 0.10 & 4.87 & 0.19 & 1.20 & 0.73 & 1.05 & 0.84\\
			
			& 2/6 & 0.88 & 0.92 & 0.96 & 0.90 & 0.97 & 0.91 & 0.97 & 0.92 & 0.96 & 1.79 & 0.54 & 1.32 & 0.72 & 1.20 & 0.73 & 1.05 & 0.84\\
			
			& 3/6 & 0.88 & 0.91 & 0.97 & 0.92 & 0.96 & 0.89 & 0.98 & 0.84 & 1.04 & 1.19 & 0.76 & 1.06 & 0.91 & 1.20 & 0.73 & 1.05 & 0.84\\
			
			& 4/6 & 0.88 & 0.85 & 1.03 & 0.94 & 0.93 & 0.95 & 0.92 & 0.88 & 0.99 & 1.15 & 0.77 & 0.99 & 0.93 & 1.20 & 0.73 & 1.05 & 0.84\\
			
			\multirow[t]{-5}{*}{\raggedright\arraybackslash \small{$(1.5,0.8,0.8)$}} & 5/6 & 0.88 & 0.90 & 0.98 & 0.93 & 0.95 & 1.02 & 0.86 & 0.93 & 0.95 & 1.07 & 0.82 & 1.02 & 0.88 & 1.20 & 0.73 & 1.05 & 0.84\\
			\cmidrule{1-19}
			& 1/6 & 0.59 & 0.58 & 1.00 & 0.59 & 1.00 & 0.60 & 0.98 & 0.59 & 0.99 & 8.53 & 0.07 & 0.77 & 0.79 & 0.91 & 0.64 & 0.63 & 0.92\\
			
			& 2/6 & 0.59 & 0.63 & 0.93 & 0.59 & 0.99 & 0.60 & 0.97 & 0.60 & 0.97 & 1.09 & 0.58 & 0.61 & 0.98 & 0.91 & 0.64 & 0.63 & 0.92\\
			
			& 3/6 & 0.59 & 0.58 & 1.02 & 0.56 & 1.04 & 0.60 & 0.98 & 0.58 & 1.02 & 0.75 & 0.80 & 0.58 & 1.04 & 0.91 & 0.64 & 0.63 & 0.92\\
			
			& 4/6 & 0.59 & 0.59 & 0.99 & 0.61 & 0.96 & 0.64 & 0.91 & 0.59 & 1.00 & 0.67 & 0.88 & 0.60 & 0.99 & 0.91 & 0.64 & 0.63 & 0.92\\
			
			\multirow[t]{-5}{*}{\raggedright\arraybackslash \small{$(1.5,0.9,0.9)$}} & 5/6 & 0.59 & 0.62 & 0.94 & 0.63 & 0.93 & 0.61 & 0.97 & 0.64 & 0.91 & 0.82 & 0.71 & 0.59 & 0.99 & 0.91 & 0.64 & 0.63 & 0.92\\
			\bottomrule
	\end{tabular}}
	\begin{tablenotes}
		\item MSE*: MSE$\times    10$; Se, sensitivity used to generate auxiliary variable $A$; Sp, specificity used to generate auxiliary variable $A$; Opt, optimal design based on full data; Two.prop, a two-wave design with proportional stratified sampling at wave 1; Two.bal, a two-wave design with balanced stratified sampling at wave 1; Single.prop, a single wave proportional stratified sampling design; Single.bal, a single wave balanced stratified sampling design.
	\end{tablenotes}
	\label{sim_1}
\end{table}

\begin{table}[H]
	\caption{Mean squared error (MSE) and empirical relative efficiency (ERE) to the optimal design for $\beta_1$ based on $1000$ Monte Carlo simulations.} 
	\centering
	\scalebox{0.9}{\begin{tabular}{llllllllllllll}
			\toprule
			\multicolumn{1}{c}{ } & \multicolumn{1}{c}{Opt} & \multicolumn{2}{c}{Prior 1} & \multicolumn{2}{c}{Prior 2} & \multicolumn{2}{c}{Prior 3} & \multicolumn{2}{c}{Prior 4} & \multicolumn{2}{c}{Two.bal} & \multicolumn{2}{c}{Single.bal} \\
			\cmidrule(l{3pt}r{3pt}){2-2} \cmidrule(l{3pt}r{3pt}){3-4} \cmidrule(l{3pt}r{3pt}){5-6} \cmidrule(l{3pt}r{3pt}){7-8} \cmidrule(l{3pt}r{3pt}){9-10} \cmidrule(l{3pt}r{3pt}){11-12} \cmidrule(l{3pt}r{3pt}){13-14}
			& MSE* & MSE* & ERE & MSE* & ERE & MSE* & ERE & MSE* & ERE & MSE* & ERE & MSE* & ERE\\
			\midrule
			1/6 & 0.16 & 0.15 & 1.03 & 0.17 & 0.92 & 0.17 & 0.92 & 0.16 & 0.96 & 0.29 & 0.56 & 0.26 & 0.60\\
			2/6 & 0.16 & 0.17 & 0.91 & 0.17 & 0.93 & 0.17 & 0.91 & 0.16 & 0.98 & 0.26 & 0.66 & 0.26 & 0.60\\
			3/6 & 0.16 & 0.17 & 0.92 & 0.17 & 0.94 & 0.16 & 1.00 & 0.16 & 0.96 & 0.25 & 0.67 & 0.26 & 0.60\\
			4/6 & 0.16 & 0.17 & 0.94 & 0.16 & 0.95 & 0.17 & 0.96 & 0.15 & 1.04 & 0.26 & 0.62 & 0.26 & 0.60\\
			5/6 & 0.16 & 0.18 & 0.89 & 0.17 & 0.94 & 0.17 & 0.90 & 0.18 & 0.90 & 0.26 & 0.62 & 0.26 & 0.60\\
			\bottomrule
	\end{tabular}}
	\begin{tablenotes}
		\item  MSE*: MSE $\times 10$; Opt, optimal design based on full data; Two.bal, a two-wave design with balanced stratified sampling at wave 1; Single.bal, a single wave balanced stratified sampling design.
	\end{tablenotes}
	\label{nwts_sim}
\end{table}

\bibliographystyle{plainnat}  


\end{document}